\documentclass[a4paper]{article}

\usepackage{icrc2013}
\usepackage[english]{babel}
\title{On the methods to determine signal attenuation curve for different surface arrays }

\shorttitle{Attenuation curves for different surface arrays}

\authors{
Jakub V\'icha$^{1}$,
Petr Tr\'avn\'i\v{c}ek$^{1}$,
Dalibor Nosek$^{2}$,
Jan Ebr$^{1}$.
}

\afiliations{
$^1$ Institute of Physics of Academy of Sciences, Prague, Czech Republic \\
$^2$ Charles University, Faculty of Mathematics and Physics, Prague, Czech Republic \\
\scriptsize{
}
}
             
\email{vicha@fzu.cz}

\abstract{Large surface arrays of current cosmic ray experiments measure the signals of electromagnetic or muonic components or their combination. The correction to the zenith angle (the attenuation curve) has to be taken into account before the signal is converted to the shower energy. Either Monte Carlo simulations or indirect estimation using collected data (Constant Intensity Cut method) can be used. However, the assumptions of composition or isotropy used for the determination of the attenuation curve can still influence the final physics results such as the energy spectrum, or modify anisotropy searches and composition analysis. Using simplified Toy Monte Carlo with an output from CORSIKA simulations we try to find several examples of what kind of effects can be caused by the methods of inferring the attenuation curve. Surface arrays of different sensitivities to electromagnetic and muonic components were considered.}


\begin{document}
\maketitle

\section{Introduction}
Large arrays of particle detectors are used for studying cosmic ray showers of ultra--high energies (higher than $10^{18}$~eV). The detected signals are sensitive to electromagnetic (EM) or muonic component or their combination. Two largest modern experiments use scintillator detectors (Telescope Array) or water Cherenkov detectors (Pierre Auger Observatory). Both experiments are located at the approximately same altitude (around 1400~m~a.s.l. equivalent to 880~g/cm$^{2}$ of atmospheric depth). Thin scintillator detectors are dominantly sensitive only to a EM component, while in water Cherenkov detectors the signal is produced by EM particles and muons as well. In any case, the signal ($S$) of the surface detector array has to be corrected for different attenuation of shower size with respect to the amount of air penetrated before reaching the detector. 

Telescope Array uses the so called look--up table from Monte Carlo (MC) simulations providing the relation between the signal size, zenith angle ($\Theta$) and the shower energy~\cite{TA:enespec}. Only proton primaries are considered. Obtained energies are then rescaled to match energies measured in the fluorescence detector. This procedure is an extension of the energy independent application of the signal attenuation curve. 

At the Pierre Auger Observatory, the so called Constant Intensity Cut (CIC) method \cite{Hersil:cicmet} is applied to the measured data \cite{AUGER:enespec} providing a relationship between the size of the signal and the $cos^2(\Theta)$ at a given intensity (energy) cut. In the next step, for each shower the signal at the reference angle ($S_{Ref}$) is calculated using normalized CIC curve. The $S_{Ref}$ value is then related to the shower energy measured by the fluorescence detector. The CIC curve is studied as a function of energy (intensity cut). Since no substantial deviations in the CIC curve shape are found, just one normalized curve is finally used for all the showers.

In this contribution we study what happens to reconstructed energies if the primary particles are of a mixed composition from protons and iron nuclei. Both possibilities - the application of the MC attenuation curve and the CIC approach are investigated separately for EM type as well as EM+$\mu$ sensitive observatory.   
Since the CIC method is based on the assumption of the uniform distribution of events in $cos^2(\Theta)$, we also tried to estimate the influence of presence of a source at the highest energies violating to some extend this uniform distribution. 

To answer these questions we use Toy MC in combination with an output from simulations produced by CORSIKA ver.~7.37~\cite{CORSIKA}. 
Based on very rough assumptions of the detector response we first calculate examples of the signal attenuation curves for proton and iron induced showers from the CORSIKA simulations. Both EM and EM+$\mu$ type observatories are assumed. These curves with an assumed size of signal fluctuations then serve as an input for Toy MC to generate a large number of events that are reconstructed by both MC-like and CIC-like approaches in the last step.  

\section{CORSIKA simulations}
Proton and iron induced showers of energy $10^{19}$~eV were chosen for simulation in 10 steps of fixed zenith angles from 0$^{\circ}$ up to 60$^{\circ}$. Most recent model QGSJet~II-04~\cite{QGSJETII04} was used to simulate hadronic interactions at high energies. The updated model EPOS-LHC~\cite{epos},~\cite{epos2} was used for comparison. As common, FLUKA model~\cite{FLUKA} was used for low energy interactions. Energy thresholds of 50 MeV and 1 MeV were chosen for muons and EM particles, respectively. In total 2400 cosmic air showers were simulated. 

Response of an EM detector (assuming thin scintillators and array density similar to Telescope Array) is supposed to be proportional to the ground density of EM particles reaching the Earth surface at 800 m from the core in the plane perpendicular to the shower axis. Signals in EM+$\mu$ type of observatory (water Cherenkov tanks and array configuration similar to the Pierre Auger Observatory) are assumed to be proportional to the muon density and the energy density of EM particles both in 1000 m from the core in the plane perpendicular to the shower axis. The absolute strengths of the muon and EM induced signals were normalized to be equal for vertical proton showers. These rough assumptions are far from the detailed understanding of detector response at current observatories and shall be taken just as two illustrative examples of EM and EM+$\mu$ type of detectors.

\begin{figure}[t]
  \centering
\includegraphics[width=0.4\textwidth]{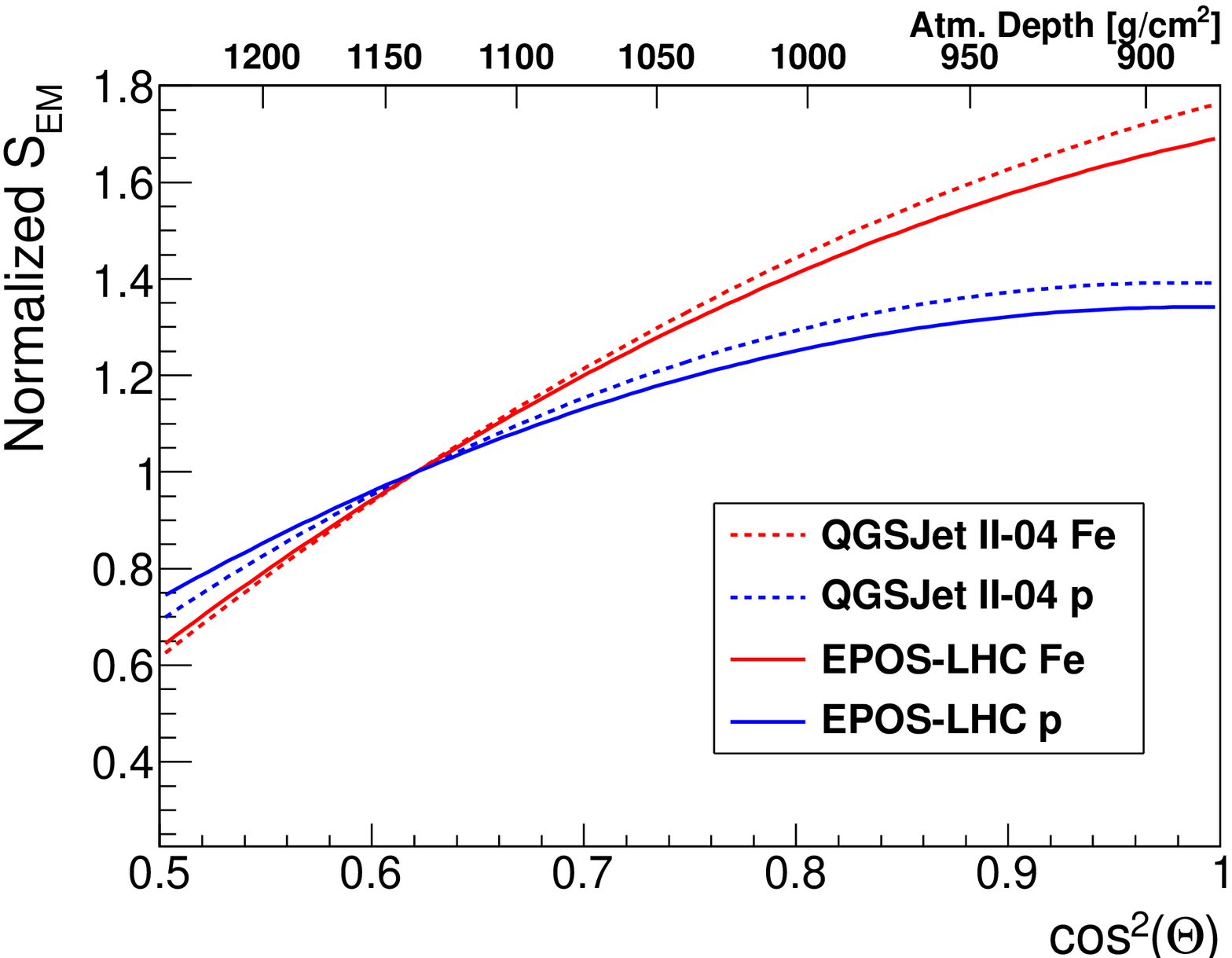}
\includegraphics[width=0.4\textwidth]{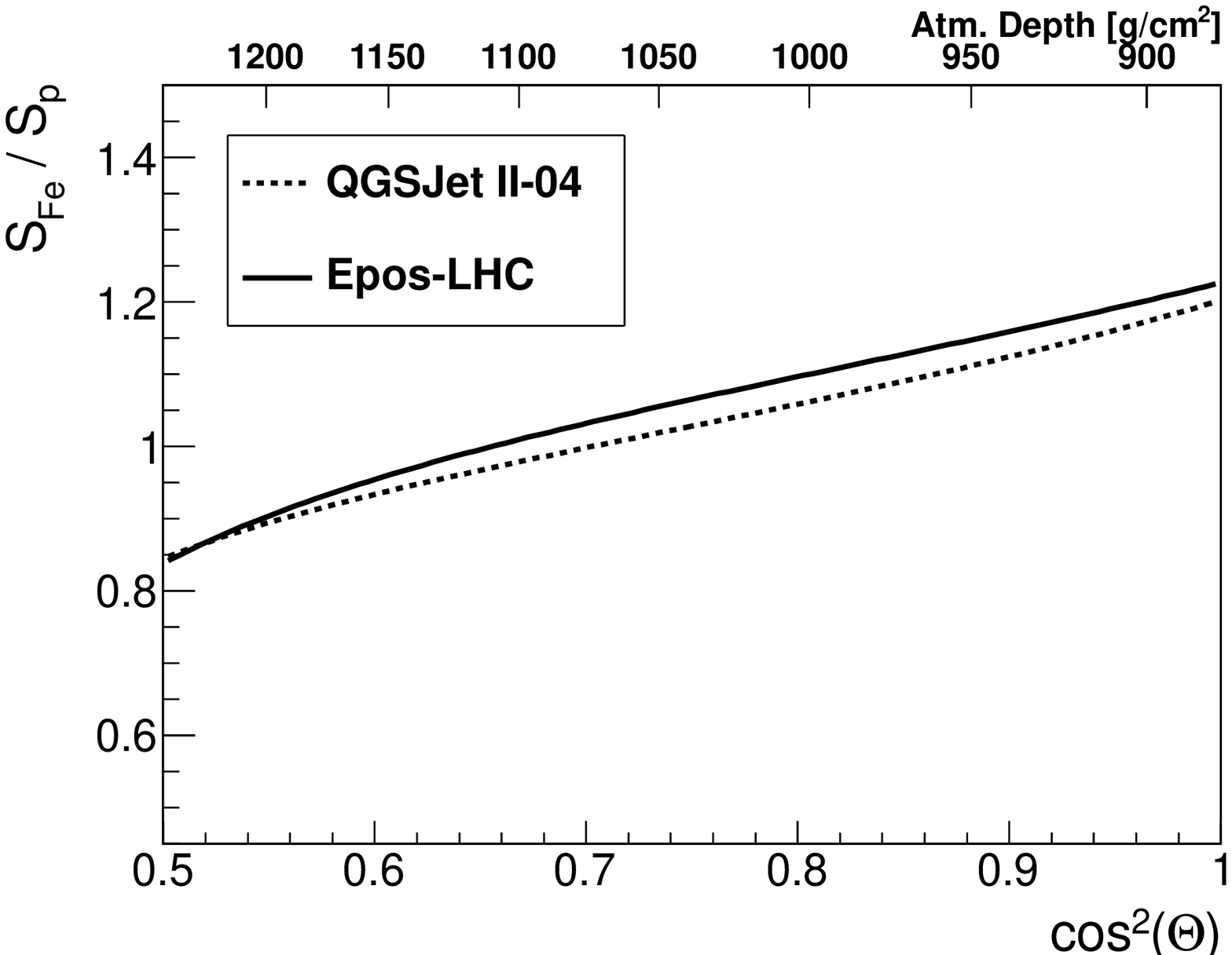}
  \caption{Attenuation of the signal normalized at 38$^\circ$ for {\bf EM} detector is shown in the top panel of figure. Ratio of absolute signals for protons and iron induced showers is illustrated in the bottom panel of figure. Corresponding atmospheric depth is shown on the upper horizontal axis. Note the smaller range of x-axis ($cos^2(0^\circ)-cos^2(45^\circ)$) wrt. Fig.~\ref{NormalizedSignalAuger}.}
  \label{NormalizedSignalTA}
 \end{figure}

\begin{figure}[t]
  \centering
\includegraphics[width=0.4\textwidth]{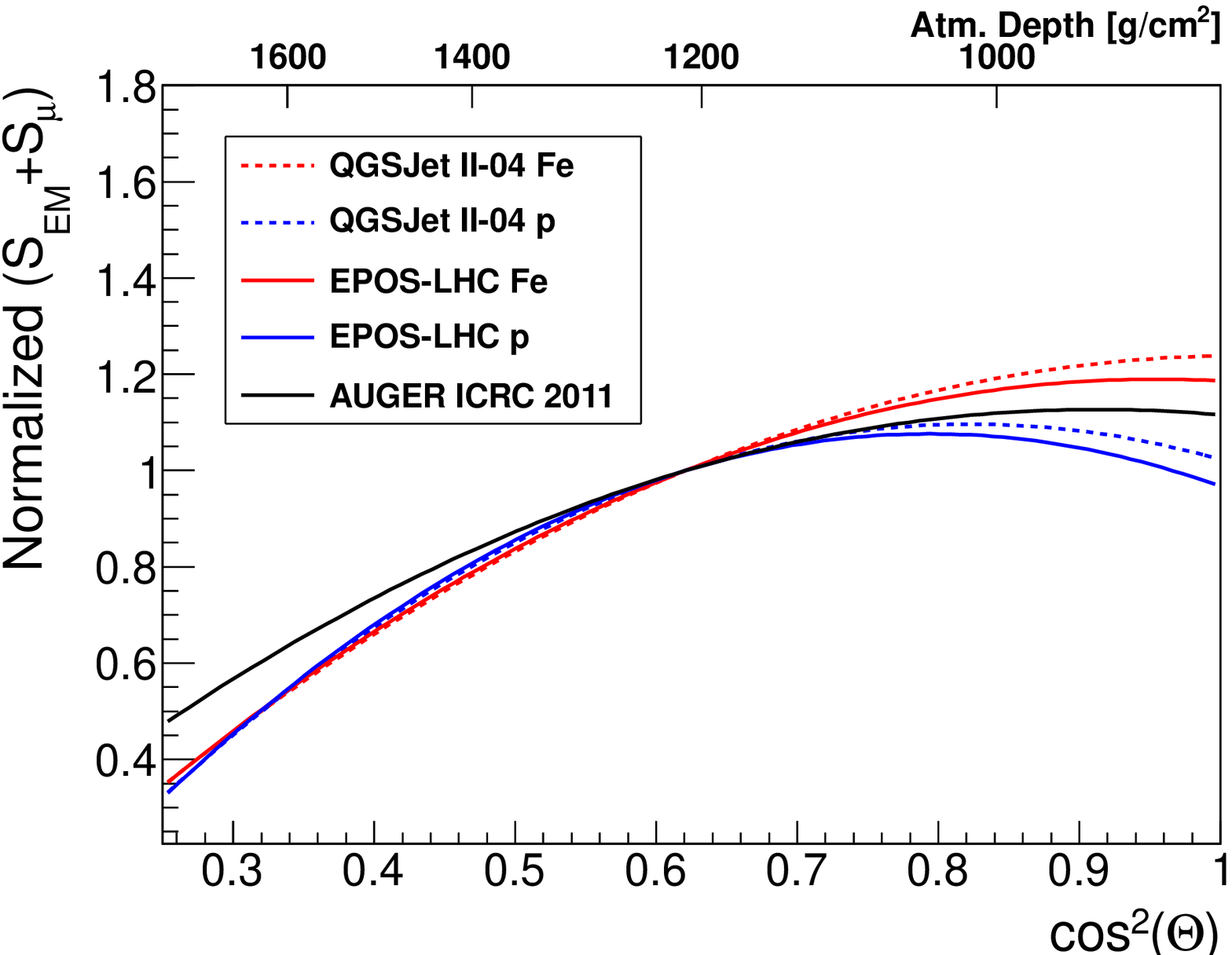}
\includegraphics[width=0.4\textwidth]{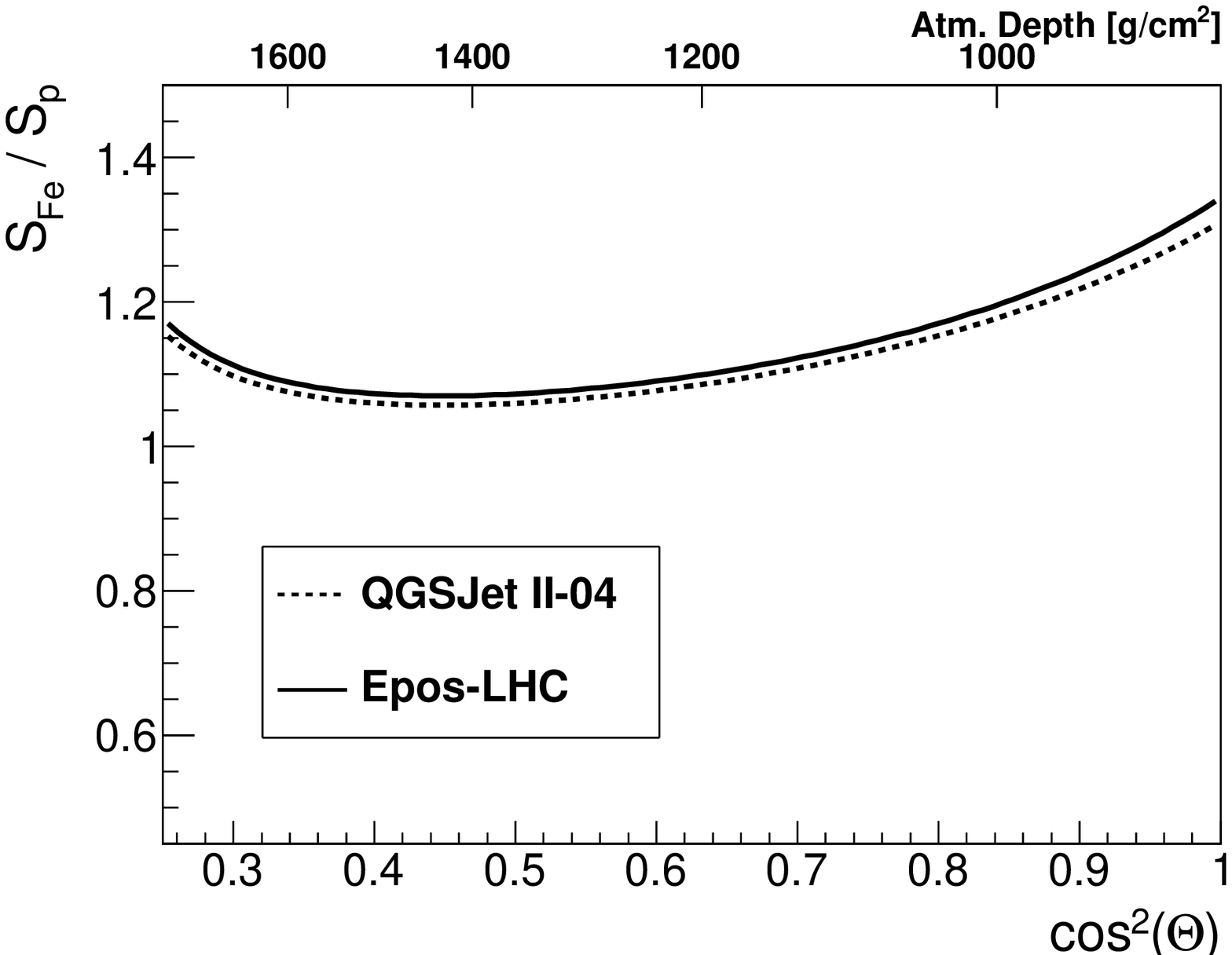}
  \caption{Attenuation of signal normalized at 38$^\circ$ for {\bf EM+$\mu$} detector is shown in the top panel of figure. Result from ICRC 2011 of the Pierre Auger Observatory is drawn for illustration. Ratio of absolute signals for protons and iron induced showers is plotted in the bottom panel. Corresponding atmospheric depth is shown on the upper horizontal axis.}
  \label{NormalizedSignalAuger}
 \end{figure}

The obtained attenuation curves are shown in Fig.~\ref{NormalizedSignalTA} for EM detector and in Fig.~\ref{NormalizedSignalAuger} for EM+$\mu$ observatory. In the upper panels, we present the attenuation curves for proton and iron induced showers. The curves are normalized at a reference zenith angle 38$^\circ$. In Fig. \ref{NormalizedSignalAuger} the published attenuation curve from the Pierre Auger Observatory~\cite{AUGER:ciccurve} derived by the CIC method is presented for completeness. In the bottom panels, the ratio of the signals induced by iron primary particles to signals from proton showers are plotted as a function of zenith angle. Both normalized shapes of attenuation curves as well as iron/proton ratio of the absolute attenuation are taken as an input for the Toy MC.    
  
\section{Toy MC}
We have simulated primary energies ($E_{MC}$) in the range (10$^{18.5}$,~10$^{20}$)~eV according to the energy spectrum 
\begin{eqnarray*}
J(E_{MC})=\frac{dN}{dE_{MC}}=E_{MC}^{\gamma}\frac{1}{1+exp^{\frac{log(E_{MC})-log(E_{1/2})}{log(W_{E})}}}
\end{eqnarray*}
i.e. by a smooth function with a steep decrease at the end corresponding to the GZK feature.
The value of $log(E_{1/2})$ was set to 19.6, $log(W_{E})$ to 0.15 and the spectral index $\gamma$ was taken as 2.7. The arrival directions were simulated isotropically (equally distributed in $cos^2(\Theta)$). Time of the cosmic ray detection was uniformly distributed within a long time period. For the studies of CIC sensitivity to presence of a source also the sky coordinates of each shower were calculated assuming the locations of the Telescope Array and the Pierre Auger Observatory. Shower energy and signal $S_{Ref}$ were assumed to be related as
 \begin{equation}
E_{MC}=a\cdot S_{Ref}^{b}
\label{eq1}
\end{equation}
 with a normalization constant $a=10^{16}$~eV. Energy exponent was taken as $b=1$ for simplicity not far from the $b$ value estimated at the Pierre Auger Observatory~\cite{AUGER:ciccurve}. This relationship in fact also reflects at least to some extend the look--up table used at Telescope Array. To get the signal $S$ for a given shower the value of $S_{Ref}$ was corrected according to the corresponding zenith angle and particle type using QGSJet~II-04 attenuation curves shown in upper panels of Figs.~\ref{NormalizedSignalTA},~\ref{NormalizedSignalAuger}. Ratios of iron and proton signals (bottom panels of Figs.~\ref{NormalizedSignalTA} and \ref{NormalizedSignalAuger}) were accounted for. Resulting signals were then smeared with Gaussian with variance equal to 10\% (5\%) of the signals for protons (iron nuclei). This way we have simulated 10$^7$ of proton and iron showers.


\section{Energy reconstruction}
 \begin{figure}[th!]
  \centering
\includegraphics[width=0.45\textwidth]{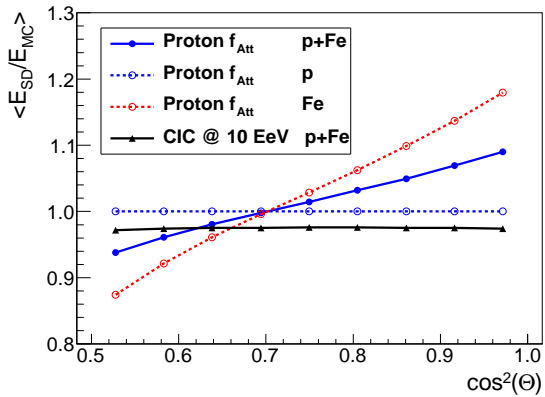}
  \caption{Ratio of reconstructed energy ($E_{SD}$) with {\bf EM} detector to MC energy ($E_{MC}$) depending on zenith angle. Proton MC attenuation curve ($f_{Att}$) is used for mixed composition 50\% p+50\% Fe, pure protons and pure iron nuclei. Results based on CIC curve~@~10~EeV are also plotted for mixed composition.}
  \label{EMZenithDependence}
 \end{figure}

After the ground signals related to the MC energy were simulated for the mixed proton/iron sample 
either the CIC method or MC--based approach were applied to obtain the reference signal $S_{ref}$ which is then transformed using Eq.~(\ref{eq1})
to get the reconstructed energy ($E_{SD}$) back.
The CIC method is based on the selection of the $N^{th}$ highest values of signal in every
bin of $cos^{2}(\Theta)$. The constant intensity cut, $N$, corresponds to the flux at certain MC energy. For MC--based reconstruction the curves corresponding to protons were chosen similarly to Telescope Array approach. 

\begin{figure}[t]
  \centering
\includegraphics[width=0.45\textwidth]{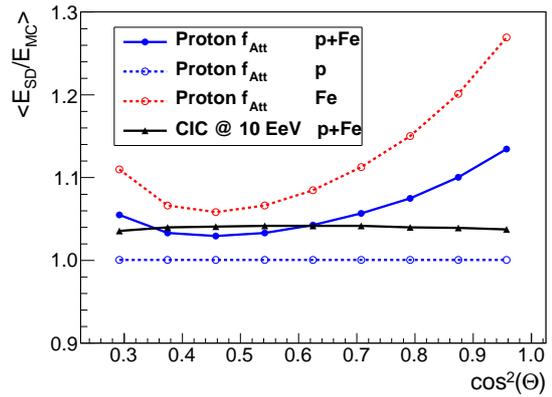}
  \caption{Same as Fig.~\ref{EMZenithDependence} for {\bf EM+$\mu$} detector.}
  \label{EMMUZenithDependence}
 \end{figure}

The performance of the methods is demonstrated in Fig.~\ref{EMZenithDependence} for EM detector and in Fig.~\ref{EMMUZenithDependence}
for EM+$\mu$ detector. The average ratio of reconstructed energy $E_{SD}$ to MC true energy, $E_{MC}$, is plotted as a function of zenith angle. Pure proton, pure iron nuclei samples as well as mixed composition, 50\%p+50\%Fe, are analyzed. A cut value, $N$, for the CIC method was chosen to correspond to around 10 EeV of MC energy. 

What should be read out from the Figs.~\ref{EMZenithDependence} and~\ref{EMMUZenithDependence} is not the absolute position of the average energy ratios, since their value is in fact always re-normalized to unity averaging the whole  $cos^{2}(\Theta)$ range when applying the calibration by the fluorescence detectors. The important behaviors are the changes of this ratio with zenith angle. 

For EM detector using the MC proton attenuation curve on pure iron sample one gets about a 30\% shift in size of $<E_{SD}/E_{MC}>$ when going from zenith angle 0$^\circ$ to $45^\circ$ (Fig.~\ref{EMZenithDependence}). For mixed composition of 50\%p+50\%Fe the shift is still about 15\%. On the other hand, the CIC approach applied to mixed composition 50\%p+50\%Fe would eliminate this zenith angle dependence to a percent level.
  
Similar conclusion can be obtained of EM+$\mu$ observatory. Fig.~\ref{EMMUZenithDependence} shows about a 25\% shift in size of $<~E_{SD}/E_{MC}~>$ when going from zenith angle 0$^\circ$ to $60^\circ$ for pure iron sample when analyzed using MC proton attenuation curve. Mixed composition with proton MC curve yields shift of 12\% while application of CIC method shows almost no zenith angle dependence. 

 Altogether, when no arguments about the composition of cosmic rays are taken as a priory assumptions (Telescope Array perhaps claims the proton like composition in the large range of energies) or if the shapes of the MC attenuation curves are not fully trusted to correspond to reality, the CIC method seems to provide more universal zenith angle description of the calibrated energies. On the other hand, possibly observed zenith angle dependence of $<E_{SD}/E_{FD}>\sim<E_{SD}/E_{MC}>$, where $E_{FD}$ denotes the energy reconstructed by a fluorescence detector, may indicate observation of mixed composition at observatories where the proton MC is used to convert registered signals to energies.

  \begin{figure}[b]
  \centering
\includegraphics[width=0.45\textwidth]{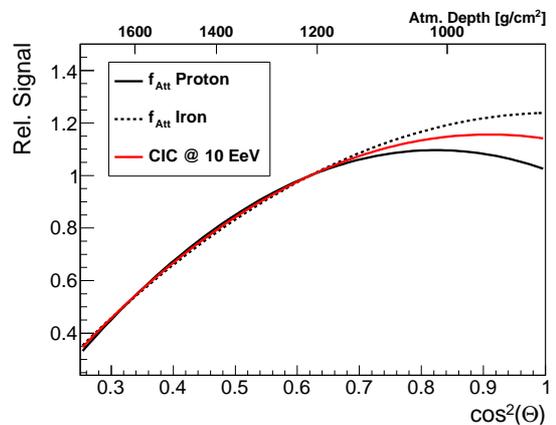}
  \caption{CIC curve~@~10~EeV and MC attenuation curves ($f_{Att}$), all normalized normalized at 38$^\circ$ for {\bf EM+$\mu$} detector. Mixed composition of 50\% protons and 50\% iron nuclei was used.}
  \label{CICEMMU}
 \end{figure}

\begin{figure}[t]
  \centering
\includegraphics[width=0.45\textwidth]{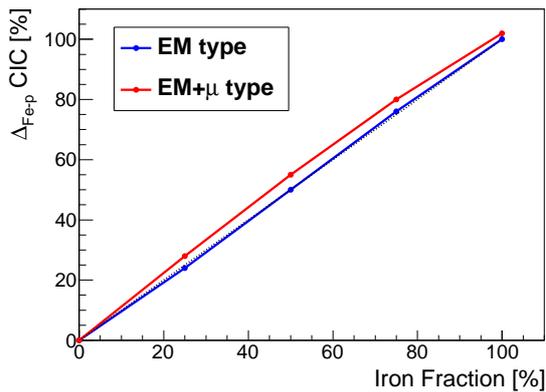}
  \caption{Relative size of CIC curve difference from MC attenuation curves depending on iron fraction. Direct proportion is shown for illustration by black dash--dotted line.}
  \label{ExperimentsCIC}
 \end{figure}

Another question we are interested in is how well the CIC method reproduces the MC attenuation curves plugged into the Toy MC. For pure protons or pure iron nuclei almost a perfect match is expected. For mixed composition the reconstructed normalized CIC curve should be in between the MC attenuation curves for proton and iron nuclei. This is demonstrated in Fig.~\ref{CICEMMU} for the EM+$\mu$ observatory and 50\%p+50\%Fe composition. Taking the normalization at 38$^\circ$ the largest difference between the reconstructed CIC and proton/iron MC attenuation curves is at $cos^{2}(\Theta)=1$. At this zenith angle we define the parameter $\Delta_{Fe-p}=D1/D2$, where $D1$ is the distance between the CIC curve and the proton attenuation curve and $D2$ is the distance between proton and iron MC attenuation curves. Parameter $\Delta_{Fe-p}$ defines the position of the reconstructed CIC curve in between the proton and iron MC attenuation curves ($\Delta_{Fe-p}=0$ - CIC is in agreement with proton MC, $\Delta_{Fe-p}=0.5$ - CIC is just in the middle between proton and iron MC, $\Delta_{Fe-p}=1$ - CIC matches the iron MC). In Fig.~\ref{ExperimentsCIC} the evolution of $\Delta_{Fe-p}$ is plotted as a function of the iron fraction both for EM and EM+$\mu$ observatories.

Perfect match to corresponding MC curves is observed for the pure iron and the pure proton showers. While $\Delta_{Fe-p}$ values obtained for EM observatory just copy the iron fraction and can be, in principle, used as a measure of the mixture, the $\Delta_{Fe-p}$ parameter for EM+$\mu$ experiment always shows higher values than the iron fraction (CIC curve stays closer to the iron MC attenuation curve). This is because the showers of the heavy component produce substantially higher ground signal. The CIC method which searches for the $N^{th}$ highest signal value in a given bin of $cos^{2}(\Theta)$ finds more likely the downward signal fluctuations corresponding to the heavy component than the upward fluctuations of the light component signals. Since the ratio of iron to proton shower signal is larger for EM+$\mu$ detector than for EM detector (bottom panels of Figs.~\ref{NormalizedSignalTA},~\ref{NormalizedSignalAuger}) the effect is present just for the EM+$\mu$ experiment.

\section{CIC method with a source} 
In case of presence of a localized source at the highest energies, the CIC method was tested by adding events with arrival directions from the 20$^{\circ}$--vicinity of Cen~A to the isotropic background of mixed composition of 50\% protons and 50\% irons. The direction of Cen~A was chosen as it is the promising region where an event excess of UHECR was found~\cite{AUGER:CenA}. We chose the energy of 50~EeV as the starting point where we add the signal to the isotropic background. The signal spectral index $\gamma$~=~2.7 was used without GZK--like suppression. Events were simulated up to the energy of 100~EeV. The CIC method is tested with EM+$\mu$ detector located in the same coordinates as the Pierre Auger Observatory. The protonic source was simulated with various strengths defined as a ratio of the number of added events to the number of isotropic events coming from the same region on the sky.

\begin{figure}[t]
  \centering
\includegraphics[width=0.45\textwidth]{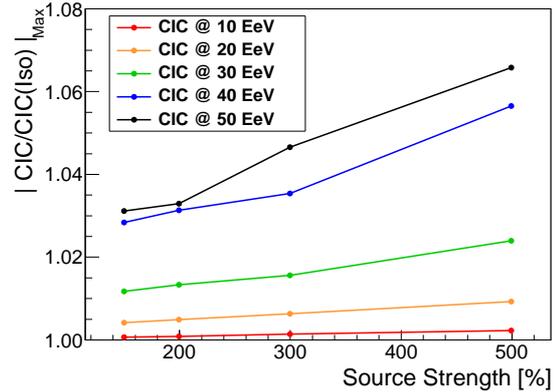}
  \caption{Maximal relative deviations of CIC curves reconstructed from isotropic plus signal events with respect to the CIC~@~10~EeV estimated from isotropic events only are plotted as a function of the source strength. Results for CIC curves at different energy cuts are indicated.}
  \label{CICSource}
 \end{figure}

In Fig.~\ref{CICSource}, the maximal deviation of the ratio of reconstructed CIC curve to CIC curve estimated at energy 10~EeV from isotropic sky is plotted as a function of the source strength. Deviations are observed to increase with the intensity cut value nearing to the threshold energy of the source and naturally also with an increase of the source strength. Still, their sizes are at a few percent level.

\section{Conclusions}
We have demonstrated that for a pure composition the CIC
method and the investigated MC--based approach are equivalent. For mixed composition, the CIC method eliminates zenith angle biases
in the energy reconstruction procedure that are present in the investigated MC--based
approach both for EM and EM+$\mu$ detectors. Once the hadronic
interaction models trustfully predict the shape of attenuation
curves, the deviation of the CIC curve from MC prediction for pure protons
could be used as a measure of composition mixture. The studied example of
source signal at the highest energies shows almost no impact on the CIC
shape for intensity cuts at low energy and small deviations for cuts close to the source energy.

\vspace*{0.5cm}
\footnotesize{{\bf Acknowledgment:~}{This work is funded by Ministry of Education, Youth and Sports of the Czech Republic under the project LG13007.}}

\end{document}